# Modeling for heterogeneous oxidative aging of polymers using coarse-grained molecular dynamics

*Takato Ishida[a*], Yuya Doi[a], Takashi Uneyama[a], Yuichi Masubuchi[a]*

[a] Department of Materials Physics, Nagoya University, Furo-cho, Chikusa, Nagoya 464-8603, Japan

\* E-mail: ishida@mp.pse.nagoya-u.ac.jp (T. Ishida)

for Table of Contents use only

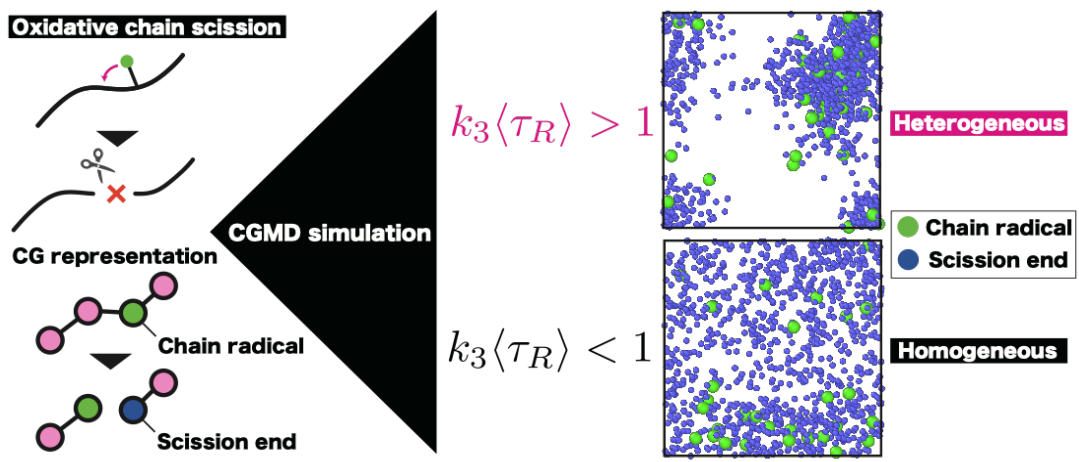

**ABSTRACT**

This study presents a coarse-grained molecular dynamics simulation model to investigate the process of oxidative aging in polymers. The chemical aging effect is attributed to the auto-oxidation mechanism, which is initiated by radicals, leading to the chain scission and crosslinking of polymer chains. In this study, we integrate a thermal oxidation kinetic model in the oxygen excess regime into the Kremer-Grest model, thereby enabling a reactive coarse-grained molecular dynamics simulation to capture the process of oxidative degradation. Our simulation reveals that when the timescale of the characteristic reaction step of oxidative degradation is shorter than the longest relaxation time of polymer chains, the scission sites exhibit spatial heterogeneity. This innovative simulation model possesses the potential to enhance our comprehension of polymer aging phenomena, thus making noteworthy contributions to the realm of polymer science and degradation chemistry.

## 1. INTRODUCTION

Oxidative aging holds significant importance in determining the long-term reliability of polymeric materials for real industrial applications. Thus, extensive research has been devoted to modeling degradation phenomena, with the aim of predicting the degradation and lifetime of such materials. One of the early approaches introduced is a microscopic chemical kinetics approach pioneered by Boland and Gee[1] in 1946. This approach utilizes a kinetic model based on the autoxidation mechanism observed in oils, adapted for application to polymers. The proposed reaction mechanism, known as the Basic Autooxidation Scheme (BAS), exhibits a noteworthy characteristic. It can effectively describe various types of chemical aging, including thermal-, photo-, or radio-aging, provided that the primary reaction mechanism involves oxidation[2–6]. The above approaches are kind of reaction kinetic models that assume a well-stirred homogeneous reaction field. While reaction kinetic approaches assuming a homogeneous reaction field have been successful, some experimental results have been presented that the homogeneous assumption may not always hold. The study conducted by Celina *et al.* [7,8] in 1995 utilized chemiluminescence imaging to visualize the aging state of thermo-oxidized polypropylene. The study revealed heterogeneous aging behavior, resembling the spread of an infectious disease. Recent experimental investigations have demonstrated that the cause of such heterogeneous degradation, analogous to the spread of infectious diseases, is attributed to the transport of radical species[9,10]. Consequently, describing systems exhibiting heterogeneous degradation solely with simple homogeneous kinetic models becomes exceedingly difficult.

In comparison to flash pyrolysis in high-temperature environments (e.g., exceeding 400 °C), oxidative aging is considered as a milder aging process that proceeds at a relatively slower rate. Notably, several full-atom MD simulations employing MD_REACT[11,12] and ReaxFF[13–15] have been reported, focusing on the simulation of decomposition behaviors under high-temperature conditions, which can be classified as flash pyrolysis. However, oxidative aging, being a gentler process than flash pyrolysis, necessitates the calculation of long-term dynamics. This poses a computational challenge when employing an all-atom model. Therefore, utilizing coarse-grained molecular dynamics (CGMD) proves advantageous in enhancing computational efficiency for oxidative aging simulations. Recently, research on reactive CGMD[16] has also advanced significantly. Several CG simulations of radical polymerization[17], polycondensation[18], and hydrolysis[19] incorporating reaction mechanisms have already been reported. Furthermore, there have been recent investigations into simulations of radio-oxidation using Monte Carlo based simulation[20] and pyrolysis simulations that combine ReaxFF with the MD acceleration algorithm[21]. These approaches aim to address the issue of computational costs associated with calculating the long-term dynamics of polymer systems involving chemical reactions.

In this study, we present a CGMD simulation of oxidative aging by integrating a reaction kinetic model of polypropylene (PP) in the Oxygen Excess Region (OER), based on the BAS, which represents the simplest case of oxidative degradation. We combine the chemical reaction kinetics with the standard Kremer-Grest model[22], a well-established CG model for polymers. The OER corresponds to a sufficiently high oxygen concentration region where the oxygen addition reaction in the BAS can be assumed to occur instantaneously[23]. Thermal oxidative aging of PP is often chosen as a model case in the field of polymer degradation, since its aging behaviors have been extensively studied experimentally, and its oxidation kinetics are well-established. The reaction kinetics of PP oxidation[24,25] are primarily governed by the hydrogen abstraction reaction by peroxy radicals[23], which corresponds to the reaction step in the BAS with a rate constant denoted as $k_3$. In this study, non-equilibrium aging simulations were performed by varying the characteristic kinetic parameter, $k_3$, in a system consisting of $N = 100$ Kremer-Grest chains. We chose to set up a system consisting of initial $N = 100$ chains and chains that undergo scission during oxidative aging, as the characteristic parameters of the chains (such as viscosity and relaxation time) are expected to follow the Rouse model[26]. Our analysis focuses on the heterogeneous structure that arises from the interplay between the reaction kinetic parameters and the relaxation time of polymer chains. We aim to elucidate the intrinsic origin of the heterogeneous aging behaviors, akin to the spreading of an infectious disease.

## 2. MODELS AND SIMULATIONS

A polymer melt consisting of 2560 Kremer-Grest chains with the bead number per chain $N = 100$, was generated in a cubic periodic box and fully equilibrated. The motion of each Kremer-Grest bead is driven by the following Langevin equation[22]

$$m\frac{d^2\boldsymbol{r}_i}{dt^2} = -\frac{dU}{d\boldsymbol{r}_i} - \Gamma\frac{d\boldsymbol{r}_i}{dt} + \boldsymbol{W}_i(t) \qquad (1)$$

where, $m$ is the bead mass, $\boldsymbol{r}_i$ is the bead position, $U$ is the total potential energy, $\Gamma$ is the bead friction, and $\boldsymbol{W}_i(t)$ corresponds the random Gaussian force obeying $\langle \boldsymbol{W}_i(t) \rangle = \mathbf{0}$ and $\langle \boldsymbol{W}_i(t)\boldsymbol{W}_j(t') \rangle = 2k_B T \Gamma \delta_{ij} \delta(t-t')\mathbf{I}$, where $\langle \cdots \rangle$ represents the statistical average and $\mathbf{I}$ is the unit tensor. Here $k_B$ is the Boltzmann constant, and $T$ is the temperature. The potential $U$ consists of $U_{\text{LJ}}$ and $U_{\text{FENE}}$. $U_{\text{LJ}}$ is the mutual interactions between all beads, and is given by the truncated Lennard-Jones (LJ) potential written as follows.

$$U_{\mathrm{LJ}}(r) = \begin{cases} 4\varepsilon\left[\left(\dfrac{\sigma}{r}\right)^{12} - \left(\dfrac{\sigma}{r}\right)^{6} + \dfrac{1}{4}\right], & r \leq 2^{1/6}\sigma \\ 0, & r > 2^{1/6}\sigma \end{cases} \qquad (2)$$

where, $r$ is the distance between two interacting beads, $\varepsilon$ is the LJ energy scale and $\sigma$ is the bead diameter. $U_{\mathrm{FENE}}$ is the FENE (Finite Extensible Nonlinear Elastic) spring potential for bonded beads given by the following form.

$$U_{\mathrm{FENE}}(r_{ij}) = \begin{cases} -0.5kR_0^2 \ln\left[1 - \left(\dfrac{r_{ij}}{R_0}\right)\right], & r_{ij} \leq R_0 \\ \infty, & r_{ij} > R_0 \end{cases} \qquad (3)$$

where $k$ is the FENE spring constant, and $R_0$ is the maximum bond length. We set the parameters to $R_0 = 1.5\sigma$, $k = 30\varepsilon/\sigma$, $k_B T = 1.0\varepsilon$, $\Gamma = 1.0$, and $\rho^* = 0.85/\sigma^3$. The bead mass $m$ was set to unity. We chose units of length, energy, and time as $\sigma$, $\varepsilon$, and $\tau = \sigma\sqrt{m/\varepsilon}$. The numerical integration was performed with the time step chosen at $\Delta t = 0.005$. We employed periodic boundary conditions, and the simulation box was a cube with a side length of 67.2. This box dimension is sufficiently larger than the averaged gyration radius $R_g = 5.26$ for the single chain with $N = 100$. All simulations in this study were conducted with the NVT ensemble.

We incorporate topological changes of chains, including chain scission and cross-linking induced by diffusing chain/free radicals. In this study, the following reactions were considered as the oxidation reaction within OER:

$$\mathrm{POOH} \rightarrow \mathrm{M}\cdot + \mathrm{OH}\cdot \qquad\qquad k_{1u} \qquad (4)$$

$$2\mathrm{POOH} \rightarrow 2\mathrm{M}\cdot + (\text{inactive product}) \qquad\qquad k_{1b} \qquad (5)$$

$$\mathrm{POOH} + \mathrm{M}\cdot \rightarrow \mathrm{M}\cdot + (\text{inactive product}) \qquad\qquad k_3 \qquad (6)$$

$$\text{POOH} + \text{OH} \cdot \rightarrow \text{M} \cdot + \text{(inactive product)} \qquad k_3 \qquad (7)$$

$$\text{M} \cdot \rightarrow \text{M} \cdot \text{(end)} + \text{(scission)} \qquad - \qquad (8)$$

$$2\text{M} \cdot \rightarrow \text{(crosslink)} \qquad - \qquad (9)$$

$$\text{M} \cdot + \text{M} \cdot \text{(end)} \rightarrow \text{(crosslink)} \qquad - \qquad (10)$$

We consider the radicals P· (polymer radical), PO· (alkoxy radical), POO· (peroxy radical), and POOH (hydroperoxide), which play important roles in BAS. They are collectively represented as the macroradical M·. The exclusion of the reaction mechanism associated with the rate constant $k_2$, which corresponds oxygen addition, is implemented based on the OER assumption. Under OER, the generated P· species promptly converts to POO·. This fast conversion allows us to represent them as M· and simplify the reaction mechanism. It is known that in the oxidation aging of PP in OER with higher oxygen concentrations than in the ambient air, the topological changes in the polymer chains are dominated by chain scission rather than crosslinking[27]. The absence of reaction rate constants for several reaction processes indicates that these processes are instantaneous reactions that occur immediately once the reaction conditions are satisfied. Figure 1 depicts a CG representation illustrating the changes in chain topology, along with the oxidation reaction considered within the BAS in this simulation. Note that we are unable to consider the reaction heat in our coarse-grained model due to its inherent limitations. However, it is worth noting that the heat generated by the reactions is not large. For instance, the bond dissociation energy of a C-C bond is approximately 365 kJ/mol[28], and the energy added per bead for each C-C bond scission is on the same order of magnitude as $k_B T$. Consequently, we assume that the heat generated by the reactions quickly reaches local equilibrium. Except for the free radicals represented by the orange beads in Figure 1, the dynamics of the other beads are governed by the equation of motion shown as eq 1. We hypothesize that the free radicals, which do not interact with other beads, exhibit significantly higher mobility compared to the rest of the beads in the system. Through the hydrogen abstraction process, these free radicals generate macroradicals on polymer chains, with the formation taking place at spatially random positions.

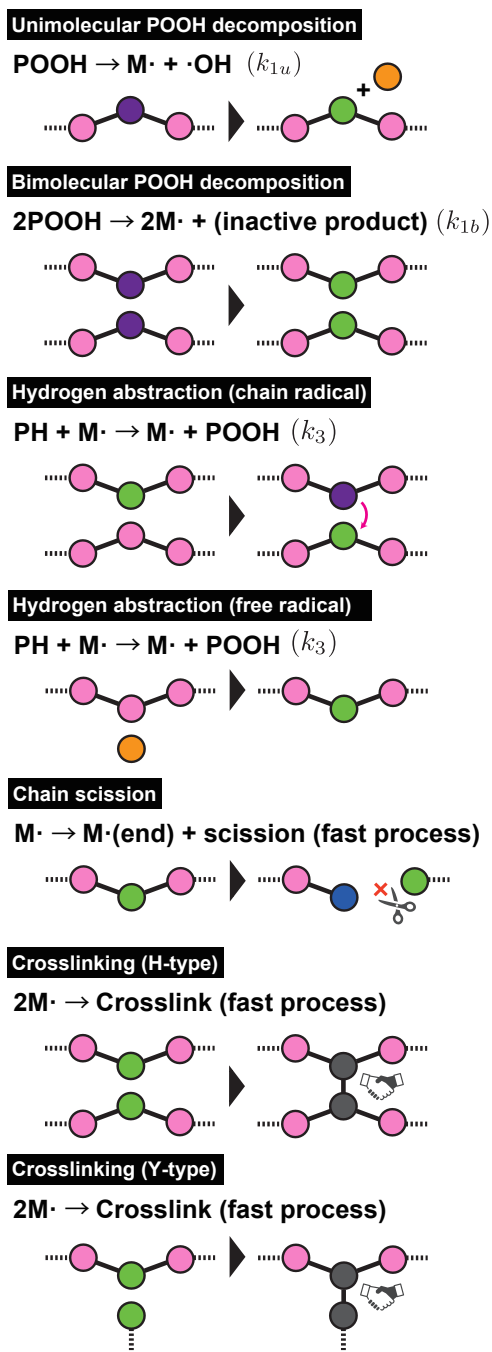

**Figure 1.** Schematic illustration of oxidation reaction paths and corresponding CG representations. Color coding is as follows. PH (polymer substrate): pink, POOH (hydroperoxide): purple, chain radical (macroradical): green, free radical: orange, scission end: blue, crosslinked beads: grey.

As previously stated in the introduction, the kinetics of oxidation in the polyolefin family, are primarily influenced by $k_3$ associated with the hydrogen abstraction reaction[23]. The selected reaction rate constants in this study are $k_{1u}/k_3 = 1.6 \times 10^{-4}$ and $k_{1b}/k_3 = 5.8 \times 10^{-3}$. These reaction parameters correspond to the thermal oxidation of PP at a temperature of 180°C in the presence of air, as referenced in the PP oxidation kinetic model developed by Richaud et al[24,25]. The "fast process" depicted in Figure 1 is a reaction mechanism that occurs significantly faster than the characteristic hydrogen abstraction reaction. In this simulation, the reaction was treated as a stochastic process based on a list of combinations that fall within the radius $r_c = 2^{1/6}$ and adhere to the spatial arrangement of the beads, as illustrated in Figure 1.

The calculations were performed using the Large-scale Atomic/Molecular Massively Parallel Simulation code, LAMMPS (version 23Jun22)[29]. For the chemical reaction calculations, the REACTION package[30] integrated within LAMMPS was utilized. The occurrence of reactions was weighted by the reaction rate constant. Initially, all beads in the system were polymer substrates, PH. Then, to initiate the simulation, one of the beads was replaced with a peroxide bead, POOH. The simulation began at the point where unimolecular decomposition (Figure 1) took place for the POOH bead. In this study, we conducted simulations to investigate the effect of varying $k_3$. We considered four distinct cases with different values of $k_3\tau_{R_0}$, where $\tau_{R_0}$ represents the Rouse relaxation time for the Kremer-Grest chain with $N = 100$ in the melt. The selected values for $k_3\tau_{R_0}$ were 60, $1.2 \times 10^2$, $2.4 \times 10^2$, and $1.2 \times 10^3$. According the previous work, $\tau_{R_0}$ is reported to be $2 \times 10^4 \tau$[31], corresponding to the longest relaxation time in the model. For example, in the case of $k_3\tau_{R_0} = 60$, the model is designed to allow the reaction to occur with a probability of $60\tau_{R_0}^{-1}$ per unit time when the beads are in the spatial arrangement that permits hydrogen abstraction.

## 3. RESULTS AND DISCUSSION

Figure 2 displays typical snapshots of the 2D-projection of the simulation box, exhibiting the positions of the scission ends and chain radicals at various conversion ratios ($\alpha = 1 - [PH](t)/[PH]_0$). Here, $[PH]_0$ represents the initial number of PH beads (i.e., 256,000), while [PH]$(t)$ denotes the remaining number of PH beads at time $t$. To maintain clarity, other types of beads are not depicted in Figure 2. In the low conversion region, an increase in $k_3$ results in the emergence of spatial heterogeneity in the distributions of both scission ends (blue dots) and chain radicals (green circles). This spatial heterogeneity is triggered by the competition between the reaction and the transportation of radicals, and it remains at higher $\alpha$.

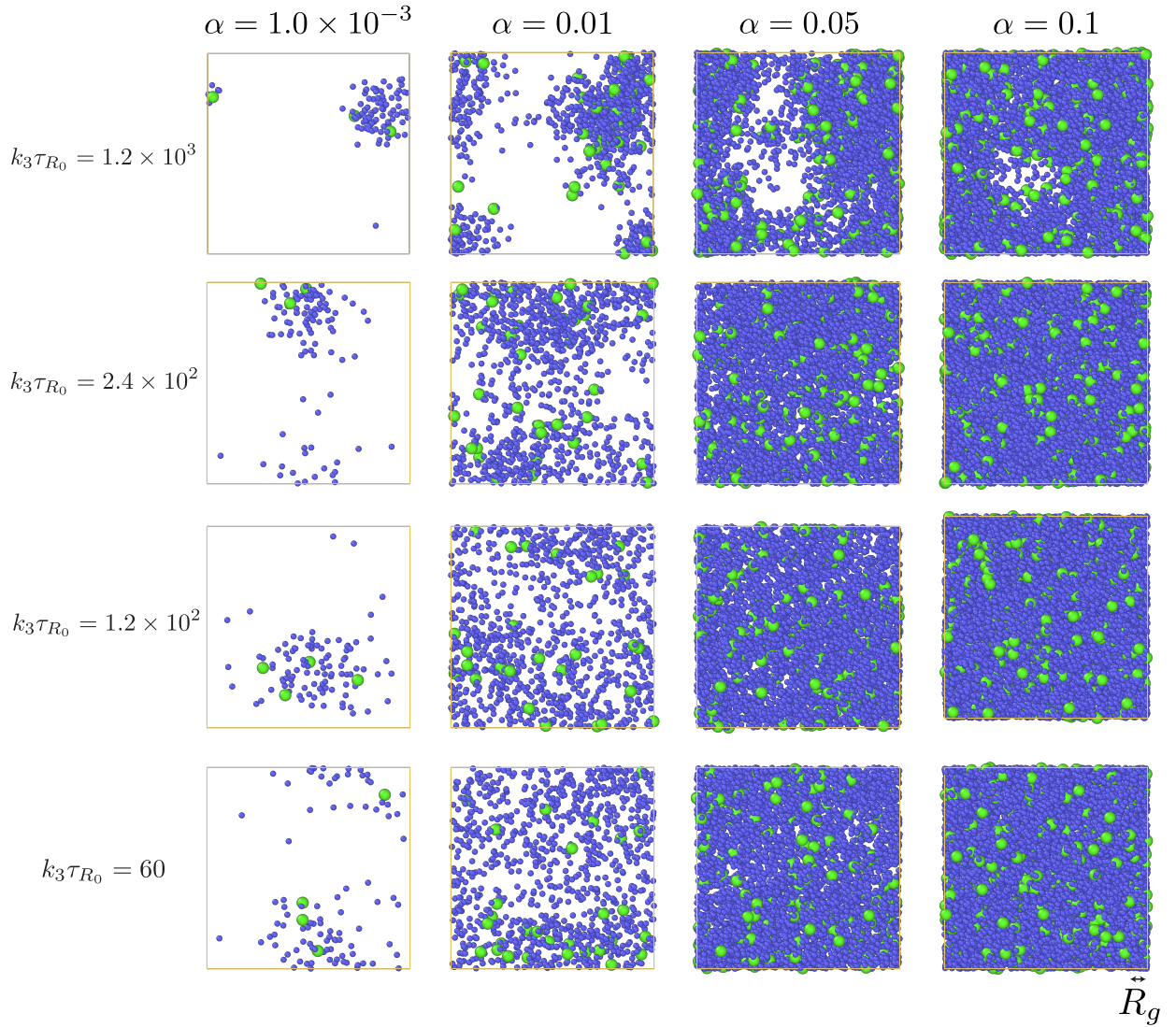

**Figure 2.** Snapshots of oxidative aging simulation for the cases of $k_3\tau_{R_0} = 60$, $1.2 \times 10^2$, $2.4 \times 10^2$, and $1.2 \times 10^3$, representing the chain radicals (M·) as green dots and scission ends as blue dots. The conversion ratio $\alpha$ corresponds to $1 - [PH](t)/[PH]_0$, and the exhibiting $R_g$ is the averaged gyration radius of the single chain with $N = 100$. For clarity, other types of beads, such as PH, POOH, OH·, are not depicted in the snapshots.

Before discussing the heterogeneity, let us look at the reaction kinetics in the system as a whole to see if the simulation is consistent with the earlier studies. Figure 3 illustrates the chemical reaction kinetics for PH consumption and the accumulation of scission ends. The presented kinetics successfully captures the auto-catalytic behavior frequently observed in the oxidation reactions of polyolefins[5,24,25,32]. As shown in Figure 4(a), $\alpha$ is proportional to $t^1$ in the induction period. The increase of chain radicals leads to the termination of the induction period (Figure 4(b)) and the subsequent auto-catalytic acceleration of the reaction dynamics. As shown in Figure 4(c), the concentration of POOH, which serves as the radical source, increases prior to the accumulation of chain radicals. This observation is in line with the BAS mechanism. Note that the impact of $k_3$ on the induction period remains uncertain. While Figure A1(in Appendix A) presents the average induction period for 16 independent simulation runs, no significant differences are observed between different conditions. In addition, the free radicals generated through the unimolecular decomposition of POOH readily undergo conversion into chain radicals through the hydrogen abstraction reaction. As a result, the number of free radicals is considerably lower than PH, POOH, and chain radicals (See Figure B1 in Appendix B). In this model, although we have incorporated crosslinking reactions to the classical BAS, the occurrence of cross-linking reactions was quite limited under the investigated conditions (Figure B2 in Appendix B). It is experimentally reported that the thermal oxidation of PP is predominantly governed by chain scission rather than crosslinking formation[27]. Figure 5 depicts the time evolution of the POOH decomposition event counts for both the unimolecular and bimolecular decomposition mechanisms. Indeed, the bimolecular process exhibits lower activation energy compared to the unimolecular process in the oxidation of polyolefin families[33]. In this system, during the auto-catalytic period, a significant quantity of POOH is present, thereby confirming the dominance of the bimolecular process over the unimolecular process under all investigated conditions (See Figure 5). The frequency of crosslinking and bimolecular POOH decomposition process may vary depending on the reaction cutoff $r_c$. However, the most characteristic reaction mechanism, i.e., hydrogen abstraction, is not significantly affected by the value of $r_c$ within the range of $1.1 < r_c < 1.3$. Therefore, the overall chemical reaction behavior is not influenced by the value of $r_c$.

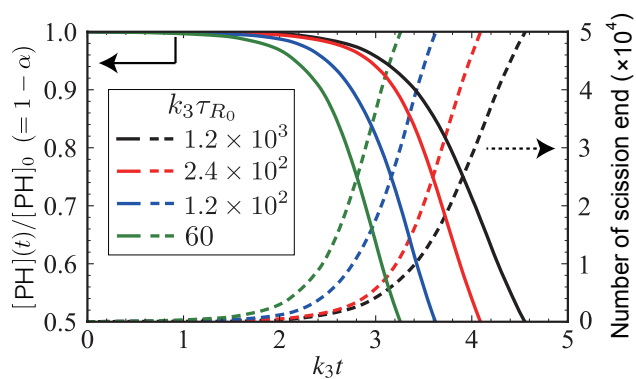

**Figure 3.** Chemical evolution during oxidative aging simulations of different $k_3 \tau_{R_0}$ for consumption of polymer substrate (PH) and accumulation of scission ends.

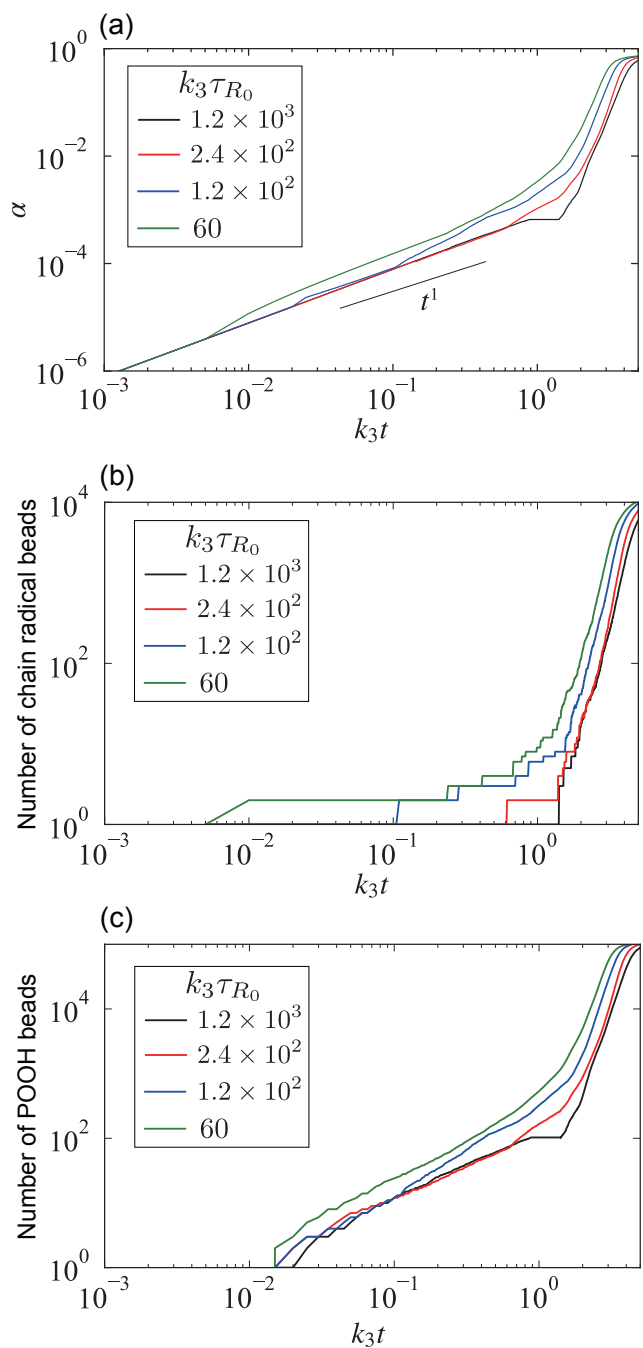

**Figure 4.** Log-log plots of various chemical reaction kinetics: (a) conversion ratio $\alpha = 1 - [PH](t)/[PH]_0$, (b) number of chain radicals, (c) number of POOH beads.

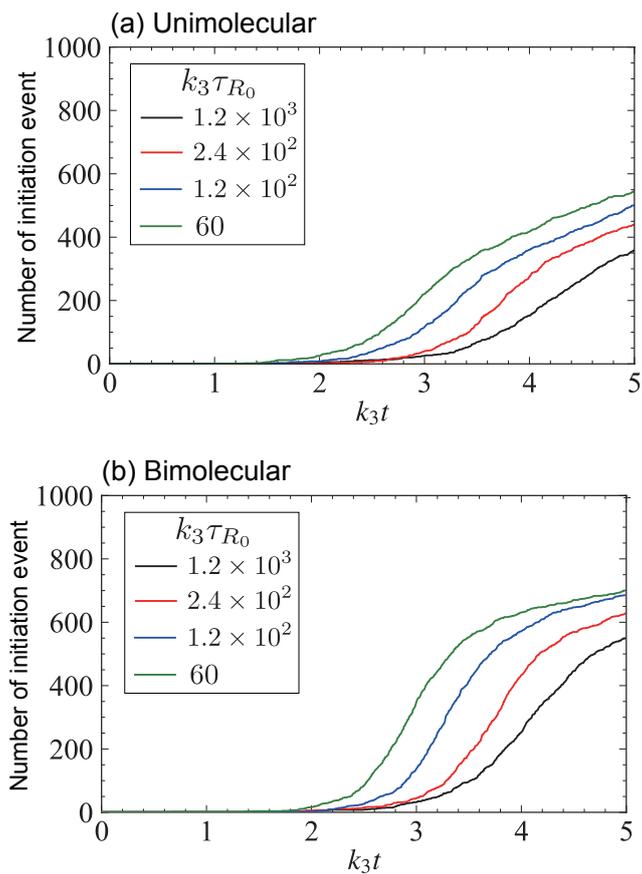

**Figure 5.** Time evolution of the POOH decomposition event counts for both the (a) unimolecular process and (b) bimolecular process.

Next, we will discuss the spatio-temporal structure of the heterogeneous aging, as observed in the snapshot of Figure 2. Figure 6 depicts the radial distribution function, $g(r)$, for all beads at various conversion ratios $\alpha$. The first peak at $r = 0.96$ corresponds to the distance between chemically bonded beads, and the second peak ($r = 1.08$) reflects the liquid ordering observed in Lennard-Jones fluids[22]. As oxidative aging progresses (i.e., as $\alpha$ increases), the first peak becomes smaller, while the second peak compensates it by becoming larger. These changes in $g(r)$ can be seen for all the examined $k_3\tau_{R_0}$, irrespective of the heterogeneity observed in Figure 2.

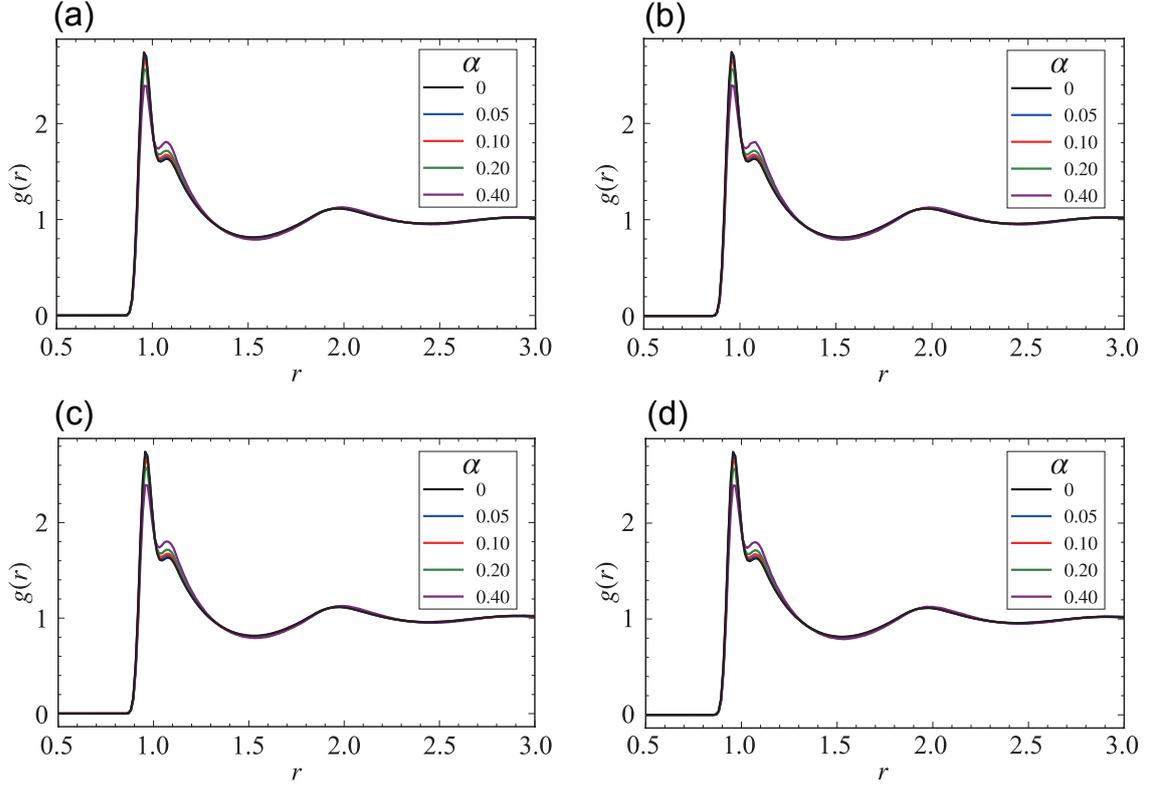

**Figure 6.** Radial distribution function $g(r)$ for all beads:

(a) $k_3\tau_{R_0} = 60$, (b) $k_3\tau_{R_0} = 1.2 \times 10^2$, (c) $k_3\tau_{R_0} = 2.4 \times 10^2$, (d) $k_3\tau_{R_0} = 1.2 \times 10^3$.

Indeed, the characteristic length scale of the spatial heterogeneity is significantly larger than the range of short-range ordering, as demonstrated by the structure factor $S(q)$ for the scission end beads (represented by blue beads in Figure 2), expressed as[34]:

$$S(q) = \left\langle \frac{1}{M} \sum_{i=1}^{N} \sum_{j=1}^{i-1} \frac{\sin qr_{ij}}{qr_{ij}} \right\rangle \quad (11)$$

where $M$ represents the number of the scission end beads in the calculation, and $r_{ij}$ represents the distance between $i$-particle and $j$-particle. In this paper, $S(q)$ is obtained as the ensemble average of 16 independent simulation runs, using snapshots at corresponding time points for each reaction rate $\alpha$. Figure 7 exhibits the structure factor $S(q)$ for various $\alpha$ ranging from 0.1 to 0.4. In almost all conditions investigated in this study, a significant increase in $S(q)$ is observed at the low-$q$ region, reflecting the heterogeneity. This result suggests that the reaction mechanism of oxidative aging based on BAS, as depicted in Figure 1, involves some intermolecular reactions, making the reaction process inherently prone to spatial heterogeneity. However, it is particularly intriguing that in the case of a low reaction rate with $k_3 \tau_{R_0} = 60$, the rise of $S(q)$ at the low-$q$ region is relatively small. As the aging progresses, this trend becomes more pronounced, and eventually, at $\alpha = 0.4$, the heterogeneity almost disappears, leading to an almost homogeneous spatial distribution.

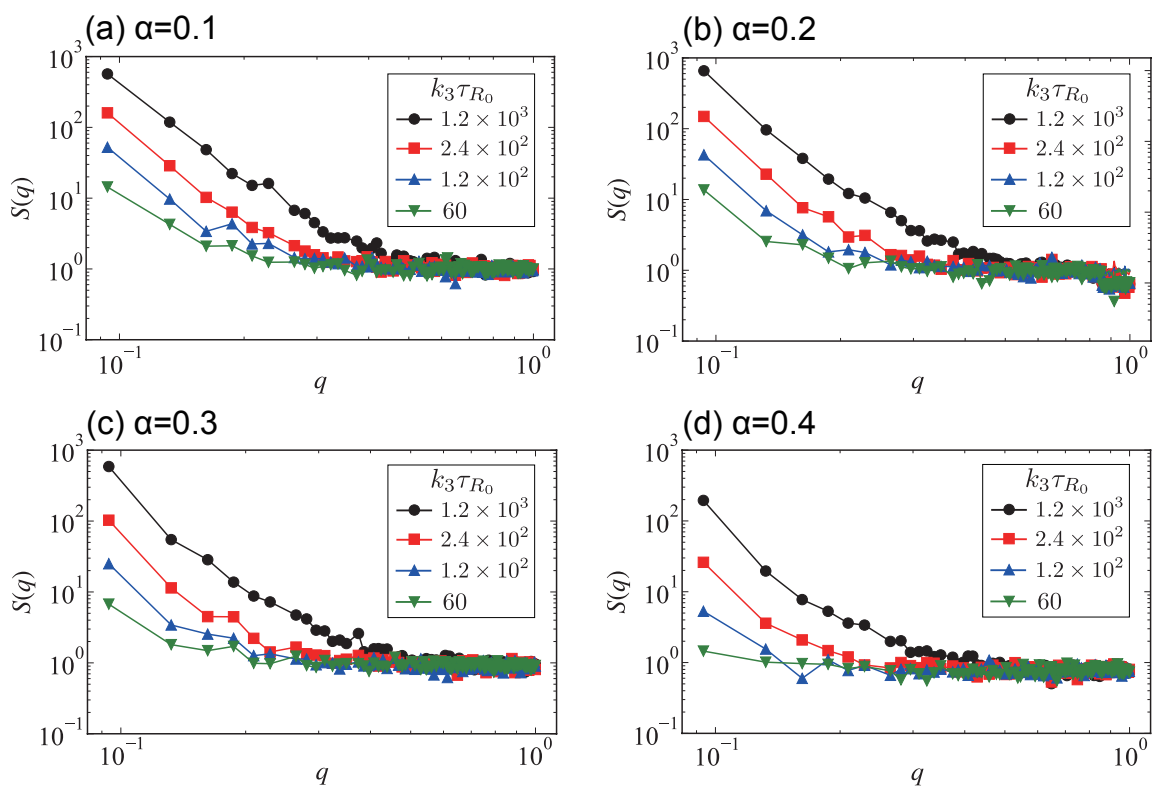

**Figure 7.** Structural factors $S(q)$ for the conversion ratio α ranging from 0.1 to 0.4:

(a) $\alpha = 0.1$, (b) $\alpha = 0.2$, (c) $\alpha = 0.3$, (d) $\alpha = 0.4$.

Let us consider the mechanism of heterogeneity in conjunction with the transport mechanism of radicals, which serve as the source of oxidation reactions. The transport mechanism of chain radicals involves a combination of polymer chain diffusion and interchain hopping facilitated by hydrogen abstraction. Therefore, in cases where $k_3$ is low or chain radicals diffuse quickly, the radicals generated on the polymer chains have sufficient time to diffuse away from their initial location before undergoing the next hydrogen abstraction reaction. It is important to note that the model used here assumes that radical initiation rate constants ($k_{1u}$ and $k_{1b}$) are proportional with $k_3$. Therefore, even with an increased $k_3$, the BAS reaction kinetics of the target system are always maintained. As a result, the system approaches a "well-stirred" condition, leading to a spatially homogeneous distribution of aged sites. The disappearance of heterogeneity in the high $\alpha$ regime, observed in the case of $k_3 \tau_{R_0} = 60$ for example, is due to increased chain mobility resulting from chain scission during oxidative aging. In contrast, in the case of the fastest reaction at $k_3 \tau_{R_0} = 1.2 \times 10^3$, the chains remain rather long and their mobility is sufficiently lower than the frequency of interchain hopping of radicals.

We envisage that the reduction of chain relaxation time associated with chain scission serves as the necessary condition for the disappearance of heterogeneity. By considering the average relaxation time $\langle \tau_R \rangle$ for degraded chains, we propose $k_3 \langle \tau_R(\alpha) \rangle < 1$ as the condition for homogeneity.

For instance, at $\alpha = 0.4$, the average chain relaxation time $\langle \tau_R \rangle$ is shortened to be less than 1/100 of that in the pristine system (see Appendix C). Thus, $k_3 \langle \tau_R \rangle$ for $k_3 \tau_{R_0} = 60$ becomes less than unity, reflecting that most of the polymer chains can relax before the characteristic reaction of hydrogen abstraction takes place, and the system becomes homogeneous. In contrast, in the case of the fastest reaction ($k_3 \tau_{R_0} = 1.2 \times 10^3$), $k_3 \langle \tau_R \rangle$ remains greater than 1 (~$10^1$) indicating that polymers do not sufficiently relax in comparison to the radical hopping, and the system remains heterogeneous.

In the high $\alpha$ regime, $\langle \tau_R \rangle$ becomes small. The diffusion of radicals becomes faster, and the radicals are well-stirred, leading to their spatially homogeneous distribution. In the extreme limit of $k_3 \langle \tau_R \rangle \to 0$, where the characteristic reaction is quite slow or the relaxation of the chains occurs very quickly, it is expected that there is sufficient transport of chain radicals between consecutive reactions, resulting in a homogeneous oxidation. A similar phenomenon has actually been reported in CG simulations of living polymerization of polystyrene[35]. When $k_3$ is large, active reaction species get trapped and localized reactions occur, whereas when $k_3$ is reduced, such behavior is no longer observed.

In relation to $\langle \tau_R \rangle$, let us discuss the evolution of the chain length distribution. The histogram depicting the distribution of chain lengths is presented in Figure 8, plotted across the range of conversion ratio α from 0 to 0.4. The open symbols displayed in the inset represent the theoretical values from the random scission model[36]. Note that the monomer component is not detected in this model because the mechanisms of further decomposition of fragments down to dimers and trimers, as well as the unzipping mechanism from the ends of polymer chains, are not considered. The qualitative behavior of the decay in the fraction of initially introduced $N = 100$ chains and the progressive increase of low molecular weight components was consistently observed in all investigated conditions in this study.

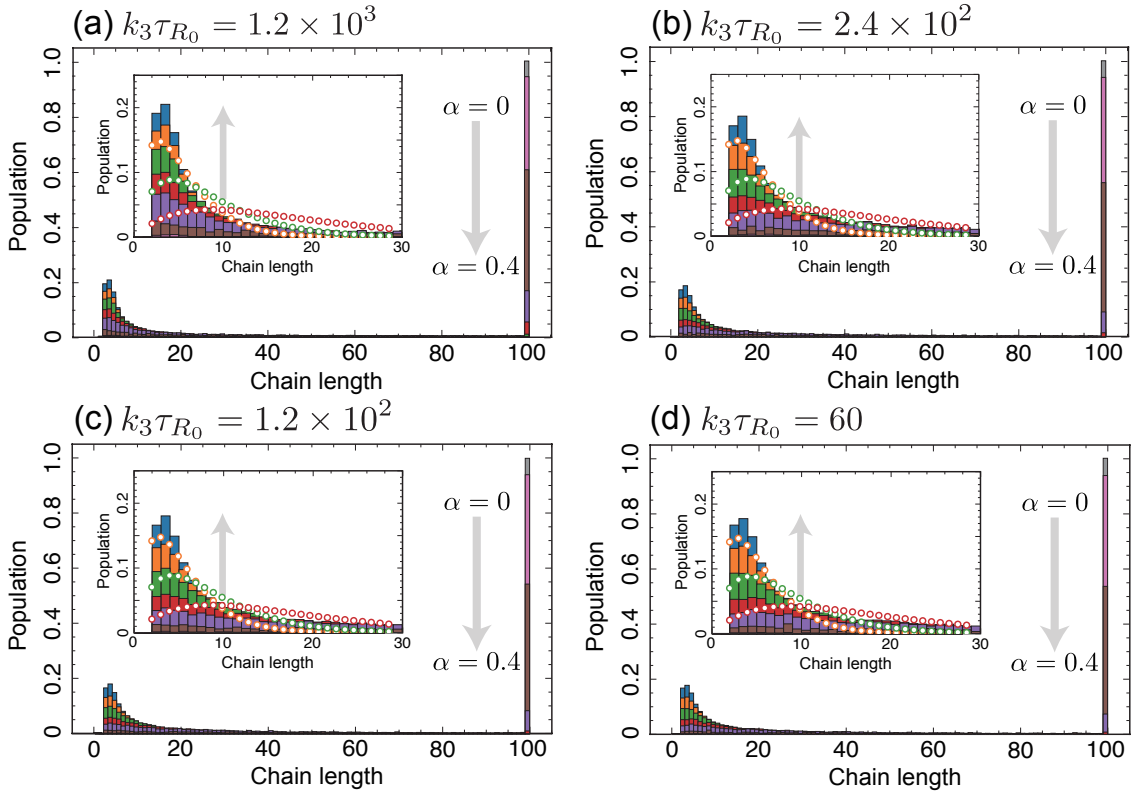

**Figure 8.** The histogram of chain lengths in the range of the conversion ratio α from 0 to 0.4: (a) $k_3 \tau_{R_0} = 60$, (b) $k_3 \tau_{R_0} = 1.2 \times 10^2$, (c) $k_3 \tau_{R_0} = 2.4 \times 10^2$, (d) $k_3 \tau_{R_0} = 1.2 \times 10^3$. The inset represents an enlarged view of the low molecular weight region.

The color coding corresponding to different values of $\alpha$ is as follows. gray for $\alpha = 0$, pink for $\alpha = 1.0 \times 10^{-3}$, brown for $\alpha = 0.01$, purple for $\alpha = 0.05$, red for $\alpha = 0.1$, green for $\alpha = 0.2$, orange for $\alpha = 0.3$, blue for $\alpha = 0.4$. The open symbols displayed in the inset represent the theoretical solution of the random scission model[36] at the conversion ratio α corresponding to the color of each histogram.

According to the findings of Jellinek and Toyoshima[37], it is known that the variation of the number average chain length $N_n$ due to random chain scission follows a conversion ratio $\alpha$ dependence of $1/N_n(\alpha) \propto \ln(1-\alpha)$. Figure 9 illustrates the relationship between $1/N_n$ and $-\ln(1-\alpha)$ for different values of $k_3 \tau_{R_0}$. Until $\alpha = 0.2$, $N_n$ align well with the model prediction[37]. Indeed, it has been previously pointed out that chain scission in the oxidative aging of PP occurs randomly[11]. The simulations conducted in this study are consistent with this observation. The deviation from the model in the high $\alpha$ regime appears to stem from the effect of notably shortened polymer chains that have become too short to undergo further scissions, thereby becoming non-negligible.

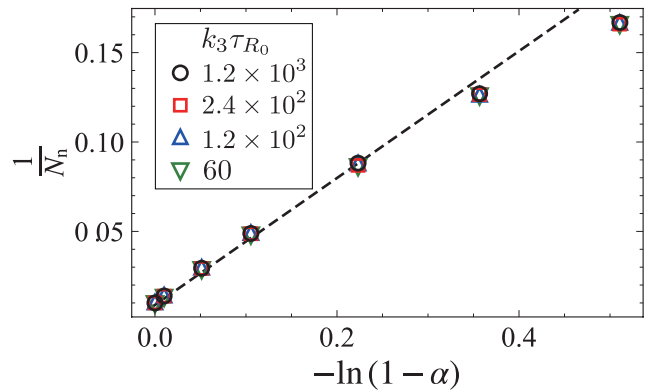

**Figure 9.** Relationship between the inverse of the number average chain length $N_n$ and $-\ln(1-\alpha)$, where $\alpha$ represents the conversion ratio. The $1/N_n$ for different $k_3 \tau_R$ values exhibit significant overlap in this plot. The dashed black line represents a linear fit based on Jellinek and Toyoshima's random scission theory[37], using data points in the range of low conversion region, $0 < \alpha < 0.1$.

It should be noted that the theoretical solution of the random scission model is not able to completely reproduce the chain length distribution in some cases. Specifically, in Figure 8(a) and Figure 8(b), where the fast reaction cases, there is an observed abundance of low molecular weight components that exceeds the prediction of the random scission model based on $\alpha$. This discrepancy can be attributed to spatial fluctuations in the local conversion ratio. In order to investigate these spatial fluctuations, we divide the simulation box into a total of $N_{\text{cell}} = 8000$ meshes, with 20 meshes along each side, and define the local conversion ratio $\alpha_{\text{local}}$ for each mesh. Then we calculate the spatial correlation function $C_\alpha$ defined as follows:

$$C_\alpha(r) = \frac{N_{\text{cell}}}{|S|} \frac{\left\langle \sum_{i,j \in S} \left( \alpha_{\text{local}}^{(i)} - \alpha \right) \left( \alpha_{\text{local}}^{(j)} - \alpha \right) \right\rangle}{\left\langle \sum_{i=1}^{N_{\text{cell}}} \left( \alpha_{\text{local}}^{(i)} - \alpha \right)^2 \right\rangle}, \quad S = \{i,j \, | \, r \leq |\mathbf{s}_i - \mathbf{s}_j| \leq r + \Delta r\} \quad (12)$$

where $\alpha_{\text{local}}^{(i)}$ represents the local conversion ratio from the i-th mesh, $\mathbf{s}_i$ is the position vector of the center of mass of the i-th mesh, and $\Delta r$ represents an infinitesimal value. We have chosen $\Delta r = 0.01$. $|S|$ denotes the cardinality of the set $S$, which serves as a correcting factor accounting for the varying number of subcells corresponding to different $r$ values.

Figure 10 demonstrates the calculated spatial correlation function $C_\alpha$ for different $k_3 \tau_{R_0}$ values. This result illustrates the spatial correlations in the difference of $\alpha_{\text{local}}$ from the global $\alpha$, and it becomes stronger as the value of $k_3 \tau_{R_0}$ increases. Furthermore, the fluctuations in the number of PH beads with each mesh ($N_{\text{PH}}^{(i)}$ from the i-th mesh) for different $k_3 \tau_{R_0}$ values are shown in Figure 11. This result also demonstrates $k_3 \tau_{R_0} = 2.4 \times 10^2$ and $1.2 \times 10^3$ exhibit larger spatial fluctuations in the local conversion ratio compared to other conditions.

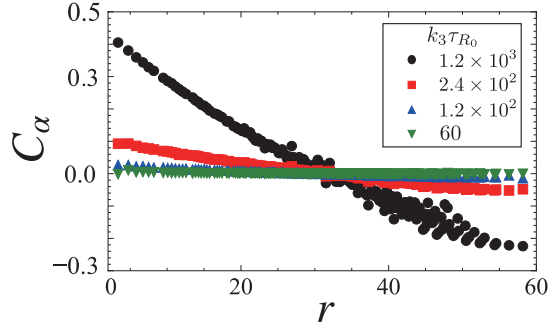

**Figure 10.** The spatial correlation function $C_\alpha(r)$ quantifies the correlation of the difference between $\alpha_{\text{local}}$ and the global $\alpha$ at different spatial distances $r$.

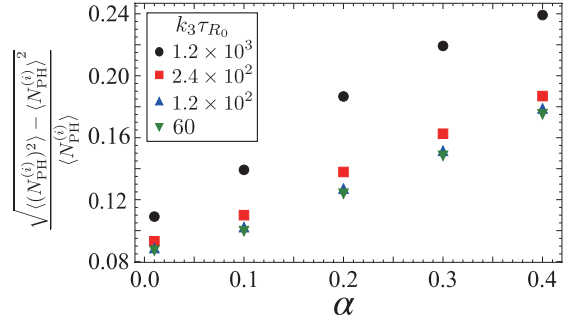

**Figure 11.** The fluctuations in the number of PH beads within each mesh ($N_{\text{PH}}^{(i)}$ from the i-th mesh).

Based on the discussion so far, we visualize the spatial distribution of short chains ($N < 10$) in our simulation, which is related to the heterogeneous progression of oxidative aging. Figure 12 presents the visualization of the spatial distribution of the beads on these short chains at $\alpha = 0.01$. In Figures 12(a) and 12(b), corresponding to $k_3 \tau_{R_0} = 2.4 \times 10^2$ and $1.2 \times 10^3$, spatial heterogeneity is seemingly observed in the distribution of short chains. Conversely, for the cases of $k_3 \tau_{R_0} = 60$ and $1.2 \times 10^2$, the distribution of short chains appears to be almost homogeneous. The snapshots shown in Figure 12 correspond to the snapshots at the same conversion ratio as the second column of Figure 2. By comparing these snapshots, we can observe that the regions with a localized scission sites and chain radicals correspond to the regions where short chains are distributed. These regions should present much higher in molecular chain mobility compared to regions where oxidative aging has not progressed significantly. For example, the diffusion coefficient of the beads within $N = 10$ chains is one order of magnitude higher than that of $N = 100$ chains on average. Therefore, it can be considered as a dynamically heterogeneous system, where there are differences in molecular mobility within the system of the fastest reaction case, $k_3\tau_{R_0} = 1.2 \times 10^3$.

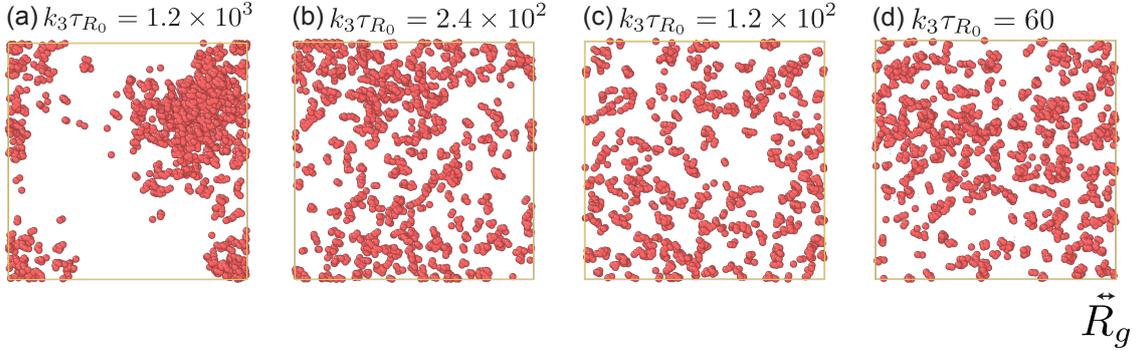

**Figure 13.** Snapshots of $N < 10$ short chain beads at $\alpha = 0.01$ in the oxidative aging simulation for the cases of $k_3\tau_{R_0} = 60,\ 1.2 \times 10^2,\ 2.4 \times 10^2$, and $1.2 \times 10^3$. $R_g$ is the averaged gyration radius of the single chain with $N = 100$.

## 4. CONCLUSION

We constructed a mesoscale simulation of polymer oxidation aging by coupling the standard Kremer-Grest model with a BAS-based chemical reaction kinetics. We generated an equilibrated melt system consisting of $N = 100$ KG chains and incorporated a reaction kinetic model for the thermal oxidation of PP at 180 °C in ambient air. The constructed model qualitatively reproduces the auto-catalytic reaction dynamics frequently observed in the homogeneous reaction kinetic model of polyolefins. It also captures the characteristic of decomposition through a bimolecular process of POOH, which is favored over a unimolecular process. The time evolution of the average chain length obtained in the simulations is consistent with the predictions of the random scission model. This agreement further supports the validity of the model used in this research and its alignment with the existing knowledge on polymer degradation.

Using the validated simulation, we discuss the spatial heterogeneity of oxidation aging, analogous to the infectious spreading, as pointed out in previous studies[10,38–40]. In summary, heterogeneous aging occurs when the characteristic reaction time is comparable to or shorter than the averaged chain relaxation time. BAS-based oxidation aging inherently introduces heterogeneity. The mobility of polymer chains is enhanced through chain scission, which can lead to the disappearance of this heterogeneous structure when the chains become sufficiently mobile. Based on the kinetic model of thermal oxidation aging for the PP used in this study at 180°C, the estimated frequency of hydrogen abstraction reactions reaches a maximum of around $10^5$ s$^{-1}$. According to Everaers *et al.*[41], if we map the relaxation time $\tau_R$ of the $N = 100$ KG chain studied in this work to real-time, it falls in the order of $10^{-7}$ s. From this study, it is evident that heterogeneous aging does not occur in the unentangled PP system due to the significant discrepancy between the charactertistic times for the reactions and the chain relaxation. The estimated timescale of hydrogen abstraction reactions from the kinetic model and the relaxation time of the $N = 100$ KG chain are separated, indicating that heterogeneity does not arise in the oxidative aging of unentangled PP. On the other hand, for example, entangled PP melt at 180 °C with a molecular weight of 20,000 is expected to have the longest relaxation time around $10^{-5}$ s. In such systems, the spatial heterogeneity of oxidation aging can become a significant issue in real applications. Therefore, conducting aging simulations specifically targeting highly entangled systems will be worthwhile for further investigation.

## ASSOCIATED CONTENT

### Author Information

### Corresponding Authors

Takato Ishida - Department of Materials Physics, Nagoya University, Furo-cho, Chikusa, Nagoya 464-8603, Japan; orcid. Org/0000-0003-3919-2348; E-mail: ishida@mp.pse.nagoya-u.ac.jp

### Authors

Yuya Doi - Department of Materials Physics, Nagoya University, Furo-cho, Chikusa, Nagoya 464-8603, Japan; orcid. Org/0000-0001-8029-7649; E-mail: ydoi@mp.pse.nagoya-u.ac.jp

Takashi Uneyama - Department of Materials Physics, Nagoya University, Furo-cho, Chikusa, Nagoya 464-8603, Japan; orcid. Org/0000-0001-6607-537X; E-mail: uneyama@mp.pse.nagoya-u.ac.jp

Yuichi Masubuchi - Department of Materials Physics, Nagoya University, Furo-cho, Chikusa, Nagoya 464-8603, Japan; orcid. Org/0000-0002-1306-3823; E-mail: mas@mp.pse.nagoya-u.ac.jp

**Notes**

The authors declare no competing financial interests.

**Acknowledgment**

This work was supported by a Grant-in-Aid for JSPS (Japan Society for the Promotion of Science) Fellows (Number 22J00300), "Nagoya University High Performance Computing Research Project for Joint Computational Science" in Japan, and CCI holdings Co., Ltd.

**Data Availability Statement**

Data analyzed during the current study are available from the corresponding author upon reasonable request.

**Appendix A: Induction period**

The time evolution of the induction period and the number of crosslinking beads for each $k_3 \tau_{R_0}$ is presented in Figure A1. It is important to note that the induction period represents the average value for 16 independent simulation runs.

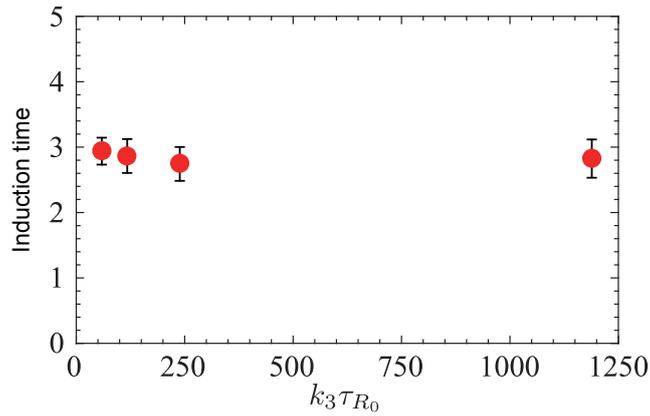

**Figure A1.** Induction periods averaged from 16 different simulation runs.

**Appendix B: Number of minor beads (free radical and crosslinked beads)**

Figure B1 represents the number of free radical and crosslinked beads corresponding to the representative result of a specific chemical reaction kinetics mentioned in the main text. The quantity of free radicals present in the system simultaneously is extremely low. This is attributed to the quick conversion of generated free radicals to chain radicals through the hydrogen abstraction process. The number of crosslinking beads is much less than the number of scission end beads shown in Figure 3, indicating that the system in this study is predominantly governed by scission processes.

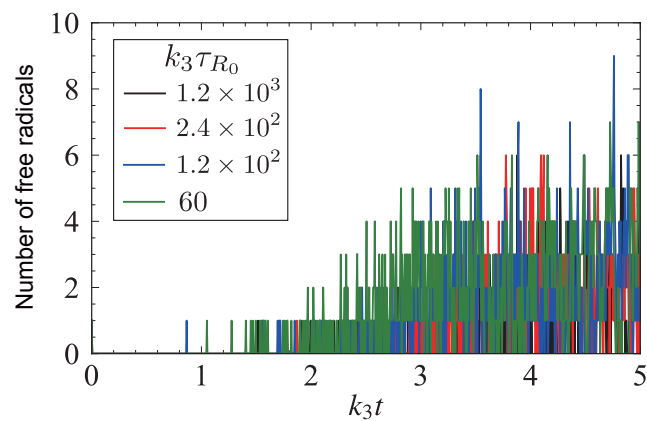

**Figure B1.** Time evolution of the number of free radicals.

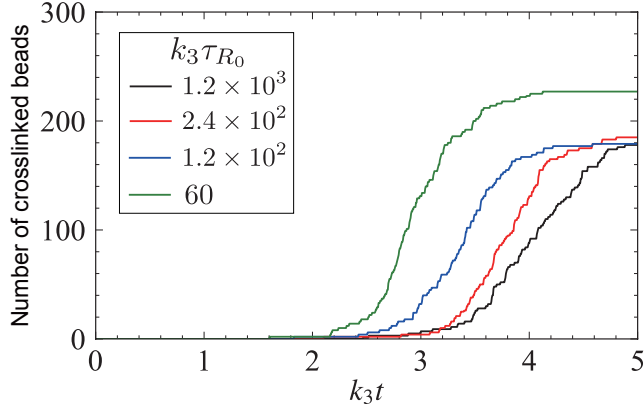

**Figure B2.** Time evolution of the number of crosslinked beads.

**Appendix C: Relaxation time**

The relaxation time of fragmented chains is anticipated to be shorter than that of the pristine polymer chains. We estimated the relaxation times of the chains after aging, according to the manner of the Rouse relaxation time $\tau_R \propto N^2$. The relationship between the average relaxation time of the systems and the oxidative aging conversion ratio $\alpha$ is depicted in Figure C1. It can be inferred that the longer relaxation time observed in the case of $k_3 \tau_{R_0} = 1.2 \times 10^3$ is due to the broader chain length distribution compared to other conditions (see Figure 8).

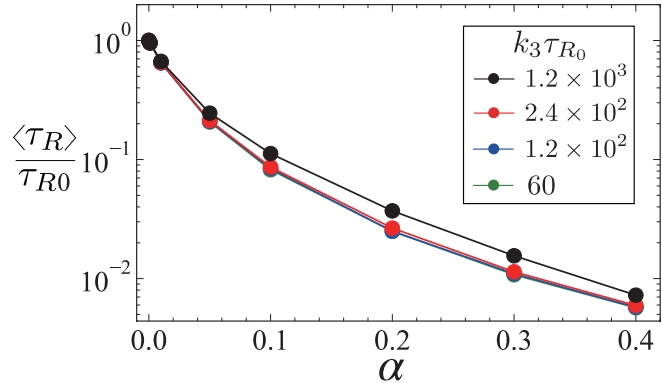

**Figure C1.** Relationship between the conversion ratio $\alpha$ and the average chain relaxation time of the aged systems.